\documentstyle[aas2pp4]{article}
\input psfig
\begin{document}

\newcommand{\greq}{\begin{equation}\left\{ \begin{array}{l}}
\newcommand{\egreq}{\end{array}\right. \end{equation}}
\newcommand{\egreqn}[1]{\end{array}\right. \label{#1}\end{equation}}
\newcommand{\beqa}{\begin{eqnarray*}}
\newcommand{\eeqa}{\end{eqnarray*}}
\newcommand{\beqan}{\begin{eqnarray}}
\newcommand{\eeqan}{\end{eqnarray}}
\newcommand{\beq}{\begin{equation}}
\newcommand{\eeq}{\end{equation}}
\newcommand{\eeqn}[1]{\label{#1}\end{equation}}
\newcommand{\beqm} {\begin{displaymath}}
\newcommand{\eeqm} {\end{displaymath}}
\newcommand{\diff}{{\rm d}}
\newcommand{\drr}{\frac{\partial}{\partial r}}
\newcommand{\dtt}{\frac{\diff}{\diff t}}
\newcommand{\dr}[1]{\frac{\partial  #1}{\partial r}}
\newcommand{\dt}[1]{\frac{\partial  #1}{\partial t}}
\newcommand{\lp}{ \left(}
\newcommand{\rp}{ \right)}
\newcommand{\lc}{ \left[}
\newcommand{\rc}{ \right]}
\def\O{\Omega}
\def\ni{\noindent}
\def\op{\Omega_p}
\def\os{\Omega_*}

\bibliographystyle{plain}

\title{Dissipation of a tide in a differentially rotating star}
\author{Suzanne Talon}
\affil{Observatoire de Paris,
Section de Meudon, 92195 Meudon, France}
\and
\author{Pawan Kumar}
\affil{
Institute for Advanced Study, Princeton, NJ 08540}

\begin{abstract}
The orbital period of the binary pulsar PSR J0045-7319, which is located
in our neighbouring galaxy the Small Magellanic Cloud (SMC), appears to be
decreasing on a timescale of half a million year. This timescale is more 
than two orders of magnitude smaller than what is expected from the standard
theory of tidal dissipation. Kumar and Quataert (1997a) proposed that this
rapid evolution can be understood provided that the neutron star's companion,
a main sequence B-star, has set up significant differential rotation. They
showed that the spin synchronization time for the B-star is similar to the
orbit circularization time, whereas the time to synchronize the surface
rotation is much shorter, and thus significant differential rotation in the
star is indeed expected. However, their calculation did not include the various
processes that can redistribute angular momentum in the star, possibly forcing 
it into solid body rotation; in that case the dissipation of the tide would
not be enhanced. The goal of this paper is to include the redistribution of
angular momentum in the B-star due to meridional circulation and shear stresses
and to calculate the resulting rotation profile as a function of time. We find
that although angular momentum redistribution is important, the B-star
continues to have sufficient differential rotation so that tidal waves are
entirely absorbed as they arrive at the surface. The mechanism proposed by
Kumar and Quataert to speed up the orbital evolution of the SMC binary pulsar
should therefore work as suggested.
\end{abstract}

\keywords{stars: binary - stars: rotation}

\section{Introduction} 
An interesting pulsar binary system was discovered in a systematic 
search for pulsars in the Small Magellanic Cloud (SMC) by McConnell et 
al.  (1991). This pulsar, PSR J0045-7319, has a spin period of 0.93 s and it
is the only known pulsar in the SMC. Its companion is a main sequence B-star 
of mass $\sim 8.8~M_\odot$ (Bell et al.  1995).  The orbital period of 
this binary system is 51 days and its eccentricity is 0.808.  
According to Kaspi et al. (1996), the orbital period of this system is 
decreasing on a time scale of $\sim 5 \times 10^5$ years, whereas the 
standard theory for tidal dissipation predicts an evolution time of
order $10^9$ years (much of our understanding of dynamical tide is due 
to a series of papers by J.-P. Zahn, cf.  Zahn 1975).  The 
problem with the standard tidal theory, as applied to a rigidly 
rotating B-star of the SMC binary, is that only a small fraction of 
the tidal energy is dissipated in one orbital period.  Kumar and 
Quataert (1997a, hereafter KQ) pointed out that if the entire tidal 
energy were to be dissipated in about one orbital period then the 
resulting evolution of the orbit would be consistent with the 
observation.  They suggested that this can be accomplished provided 
that the rotation in the interior of the B-star deviates significantly 
from the rigid body rotation.  This is in fact what is predicted by 
the work of Goldreich and Nicholson (1989) who showed that  
the angular momentum carried by tidal waves is deposited first 
near the surface of the star where the waves are dissipated. Thus, the
surface can be brought into synchronous rotation on a time 
scale short compared to the time it takes for the star as a whole 
to become synchronous. The time scale for the latter process for the 
SMC pulsar binary system is shown to be comparable to the orbital 
circularization time (Lai 1996, Kumar and Quataert 1997a) and thus the 
star is not expected to be rotating synchronously; the observed 
apsidal motion of the star provides support for this picture (see Lai 
et al.  1995, Kaspi et al.  1996). However, if angular momentum is
efficiently redistributed in stellar interiors, then a star can continue
to rotate rigidly inspite of the tidal torque operating at the surface.
The goal of this paper is to access the efficiency of angular momentum
redistribution and to calculate the expected magnitude of differential
rotation for the B-star of the SMC binary system.
In \S2, we describe our models and the physical
mechanisms considered in this study.
The main results are summarized in \S3.

\section{Angular velocity profile of a star due to
tidal torque, meridional circulation and shear stresses}
We consider the evolution of the rotation profile of a main sequence B-star 
of 9 $M_{\odot}$. We consider the star to be rotating uniformly until 
such a time so that the radius of the B-star, as a result of the normal 
stellar evolution, becomes $\sim 6~R_\odot$. Fig. 1 shows the radius of 
the star and the size of its convective core as function of time.

The energy in the dynamical tide depends on the rotation rate of star. 
The tidal 
energy in a B-star, for parameters corresponding to the SMC binary system, 
and angular rotation speed perpendicular to the orbital plane of $-1.6\times
10^{-5}$ rad s$^{-1}$ ($6 \times 10^{-6}$ rad s$^{-1}$), is 
about $6 \times 10^{40}$ erg ($10^{40}$ erg) (see KQ); the minus sign
refers to retrograde rotation. The average angular momentum 
luminosity associated with the tidal gravity wave for these two cases 
are $3 \times 10^{38}$ gm cm$^{2}$ s$^{-2}$ 
($10^{38}$ gm cm$^{2}$ s$^{-2}$) respectively.  

The dominant process for tidal wave absorption in early type stars is the
radiative damping which arises when photons diffuse from regions that are
compressed to regions that are expanding. Since the photon mean free path 
increases rapidly with distance from the center of the star, most of 
the dissipation of the wave occurs close to the stellar surface.
The fraction of wave momentum luminosity absorbed can be
calculated using $f_{\rm damp}=\exp \lc \tau (r_u) \rc $ where
\beq
\tau(\omega, \ell, r) = [\ell(\ell+1)]^{3\over2} \int_{r_l}^r K  \; {N N_t^2 \over
\omega^4}  \left({N^2 \over N^2 - \omega^2}\right)^{1 \over 2} {\diff r
\over r^3},
\eeqn{integral}
$\ell=2$ for the quadrupole tide, $K=c/\rho\kappa$ is the thermal diffusivity,
$r_l$ is the lower turning point defined by $N(r_l)=\omega$, $r_u$ is the upper
turning point defined by $r_u = 6^{\frac{1}{2}} c_s / \omega$ ($r_l<r_u$),
$c_s$ is the sound speed, $c$ is the speed of light, and $N$ is the 
Brunt-V\"ais\"al\"a frequency which can be written as a sum of two parts,
one of which is depends to the thermal gradient ($\nabla$) and the other 
on the gradient of the mean molecular weight ($\nabla_\mu$) i.e.

\beq
N^2 = N_t^2 + N_\mu^2 =
{g \delta \over H_P} (\nabla_{\rm ad} - \nabla) +
{g \varphi \over H_P}  \nabla_\mu
\eeq
(see Unno et al. 1989, or Zahn et al. 1997 for details).
The local luminosity of the wave is then given by
\beq
{\cal L}_J(\omega, \ell, r) = {\cal L}_J(\omega, \ell, r_l)\,  
\exp[ -\tau(\omega, \ell, r)].
\eeq

If the star is rotating as a solid body, only the waves
of frequency less than about $1.2 \times 10^{-5}$ rad s$^{-1}$ 
are completely absorbed as they reach the surface.  For higher frequencies,
the fraction of the wave flux absorbed decreases as $\sim \omega
^{-5}$. However, differential rotation modifies that picture.
Then, the frequency for the quadrupole tidal wave in the local rest frame of
the fluid is $\omega (r) = \omega_{t}-2\Omega (r)$, where $\Omega (r)$ is the 
angular speed of the star, $\omega_{t}\approx 2\Omega_{p}$ is the tidal 
frequency as seen in the inertial frame, and $\Omega_{p}$ is the orbital 
angular speed of the star at periastron. In a differentially rotating star,
the local tidal wave frequency decreases outward and the wave dissipation rises
rapidly. As waves deposit their angular momentum near the surface, the 
``synchronized front'' starts at the surface and progresses inwards.
However, one must also consider the role of
the redistribution of angular momentum
throughout the star as a result of meridional circulation 
and shear stress. These processes are 
calculated using the procedure described in Talon et al. (1997).

The evolution of the angular momentum in the model is governed by
\beqan
\rho \dtt \lc r^2 {\O}\rc &=& \frac{1}{5 r^2} \drr \lc \rho r^4 {\O}
U \rc + \frac{1}{ r^2} \drr \lc \rho \nu_v r^4 \dr{\O} \rc \nonumber \\
&&- \frac{3}{8\pi} \frac{1}{r^2} \drr{{\cal L}_J(r)}
\label{ev_omega}
\eeqan
where $\rho$ is the density,
$\nu_v$ is the vertical (turbulent) viscosity and
$U$ is the amplitude of the vertical circulation speed.
The latter is calculated considering differential rotation in radius,
as described in Zahn (1992) and in Matias et al. (1997),
neglecting the role of the mean molecular weight gradients.
The gradient of mean molecular weight would somewhat increase the 
timescale for the redistribution of momentum by the meridional 
circulation in the central part of the star.
However, it would influence only marginally the results for the
outer portion, which is the main object of this paper.
The turbulent viscosity takes into account the weakening effect
of the thermal diffusivity as first described by Townsend (1958).
The details concerning that viscosity are not important here
since the circulation, being an advective process, dominates the transport
of momentum almost everywhere.

The local time scale that determines the redistribution of angular momentum
due to meridional circulation is the Eddington-Sweet time scale. Fig. 2
shows this time scale as a function of $r$ in the radiative part of
a 9$M_\odot$ main sequence star of age 19.7 Myrs. 
Since the Eddington-Sweet timescale is much shorter in the
outer portion of the star, we expect the outer region to be rotating
almost as a solid body whereas differential rotation can persist
in the inner part\footnote{Actually, meridionnal circulation does not
tend to produce solid body rotation but rather a state of mild differential
rotation with $\Omega_{\rm core}/\Omega_{\rm surface} \sim 1.2$ for a 
9 $M_{\odot}$ star (see Talon et al. 1997 for more details).}.

Equation (\ref{ev_omega}) is solved to determine the rotation profile as a 
function of time. For the model with mild retrograde solid body rotation,
the upper turning point of the tidal wave is located at a radius of 
0.85$R_*$ and only about 1.0 \% of its angular momentum luminosity is deposited
during one orbital period. In this case, meridional circulation is efficient
in redistributing the angular momentum in the interior of the star.
In fact, as long as less than $\sim$ 30 \% of the angular momentum luminosity of 
the tidal wave is deposited in the star, meridional circulation carries 
away almost all of this angular momentum and spreads it over much of the 
star, and thus, little differential rotation can be maintained.
When wave damping is inefficient, the result of the rotation 
evolution calculation is dependent on the uncertain initial 
rotation profile of the star, and the time dependence of the 
angular momentum luminosity of the tide.

For these reasons we have carried out a restricted evolution calculation. We
start with a state of differential rotation in the star so that a substantial 
fraction of the angular momentum luminosity of the tidal wave is deposited 
in the outer part of the star. We then calculate how the meridional 
circulation, 
and shear turbulence modify the rotation profile. In particular we wish
to find an initial rotation profile leading to an increase of the differential
rotation;
a decrease in the differential rotation would decrease the
rate of energy and angular momentum deposit and thus increase the time scale
for the orbital evolution. To this end our calculation
was started with the rotation profile shown in Fig. 3a (at 20 Myrs). 
In that case, the rate of angular momentum deposit was about 50 \% of the 
luminosity. The initial rotation profile for prograde rotation was solid-body,
as shown in Fig. 3b; in this case, about 80 \%
of the angular momentum luminosity of the wave is deposited 
near the upper turning point (at $\sim$ 0.96 R$_*$). 
The subsequent evolution of the rotation profile in each of these two
cases is shown in Fig. 3; note that the difference between the rotation
rate in the core and the surface is increasing with time until the surface is
nearly synchronized.
In Fig. 4, we show the wave damping as a function of $r$ for every
rotation profile shown in Fig. 3. Note that in all of these cases
the tidal wave energy is completely dissipated as the wave approaches the
stellar surface, and thus the angular momentum transported by the meridional 
circulation and shear induced turbulence does not modify the rate of
change of the orbital energy for the SMC binary pulsar system PSR J0045-7319
estimated by Kumar and Quataert (1997). Thus, the time scale for the
evolution of the orbital period calculated by these authors is unaffected
by the redistribution of angular momentum in the star.

We have also calculated the effect of the differential rotation induced
turbulence on the tidal wave damping. We find that it can be an
important contributor to the wave damping at initial stages of evolution,
when the surface of the star is rotating differentially, helping to enhance 
the fraction of tidal angular momentum luminosity deposited in the star.

We would like to point out one simplification made in our calculation. 
Even though the energy in the tidal waves is modified as the rotation 
profile of the star evolves we took it to be constant. However, the error 
made is not large since the tidal waves are mostly excited near the boundary
of the convective core and radiative envelope where the rotation is nearly constant.

We note that the contribution to the wave damping from the region near the
bottom of the radiative zone is negligible inspite of the fact that the
mean molecular weight has a large gradient there leading to a sharp
increase of the Brunt-V\"ais\"al\"a~frequency. The reason is simply that
the photon mean free path here is also very short and it takes a long time
for photons to diffuse across a wavelength.
Even if 
the dissipation in that region is about 5 times greater than just outside
of it, the contribution to the total damping remains negligible.

\section{Conclusion}
We have calculated the evolution of the rotation profile of a 9 
$M_{\odot}$ main 
sequence star which is subject to a tidal torque from its companion 
neutron star and in which angular momentum is redistributed by meridional 
circulation and shear stresses.  
We calculated models corresponding to two different rotation states:
one model has a retrograde initial rotation of $-1.6\times
10^{-5}$ rad s$^{-1}$, and the other has a prograde initial rotation
of $6 \times 10^{-6}$ rad s$^{-1}$.
The magnitude of the tidal torque was taken to be 
$3 \times 10^{38}$ gm cm$^{2}$ s$^{-2}$ for the retrograde rotation
and $10^{38}$ gm cm$^{2}$ s$^{-2}$ for the prograde rotation.
The location of the dissipation of the momemtum luminosity was calculated
as a function of the rotation profile.
The meridional circulation and shear stresses were 
calculated following the method described in Talon et al. (1997).  We 
find that the fractional difference between the surface and the 
interior rotation rate in the star is a factor of a few. 
With this differential rotation, the frequency of tidal waves in the 
rest frame of the star is low enough that they are completely absorbed 
before they reach the upper turning point. For retrograde rotation of 
frequency about $6\times 10^{-6}$ rad s$^{-1}$ or greater and a significant 
amount of differential rotation, the energy dissipated per 
orbit is about 10$^{41}$ erg, which leads to orbit evolution time 
of $\sim 5\times 10^5$ years, 
consistent with the observations.

\begin{figure}     
\centerline{
\psfig{figure=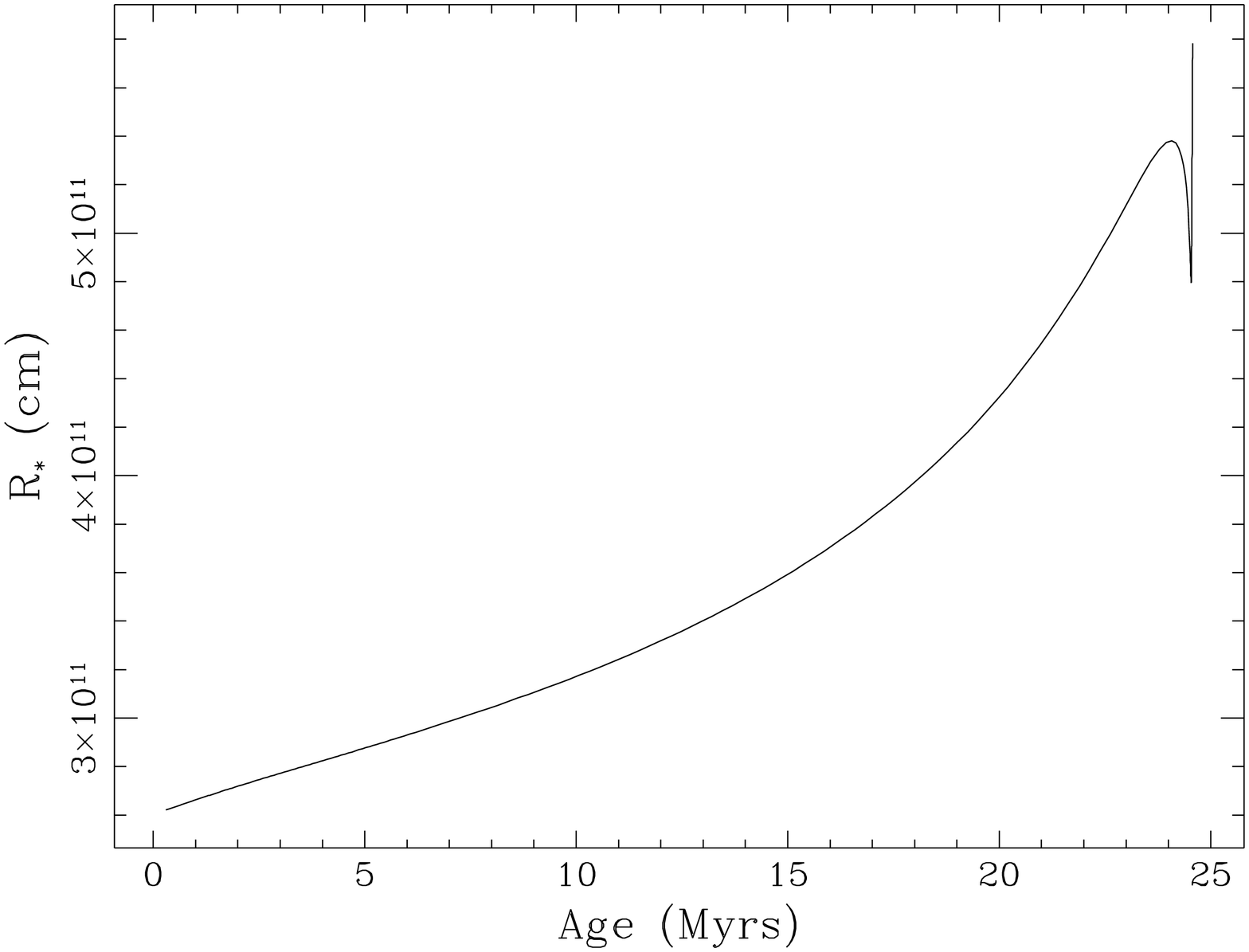,height=6cm,angle=-90}
}
\centerline{
\psfig{figure=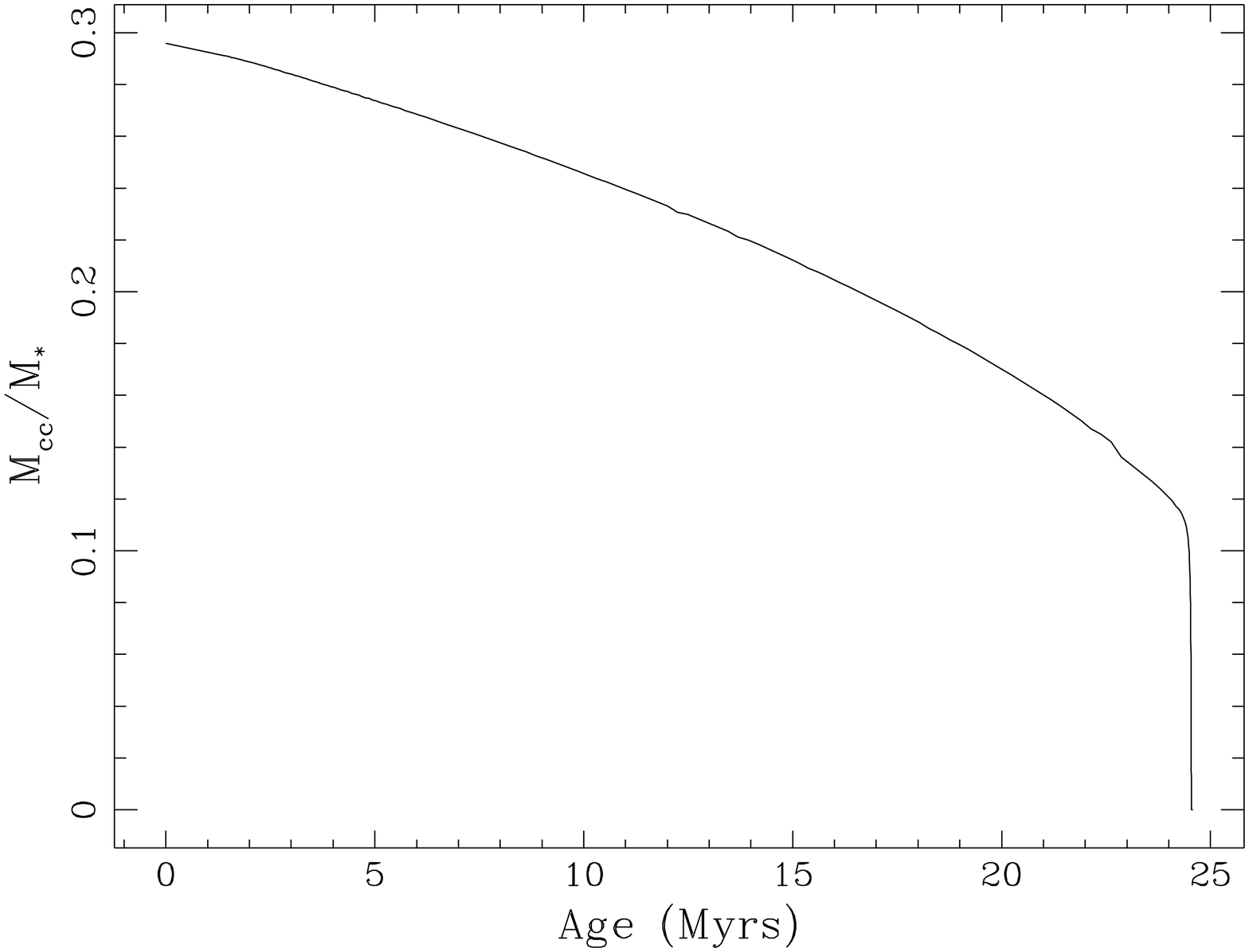,height=6cm,angle=-90}
}
\caption[]{
The radius of the B-star as a function of time is shown
in the top panel, and the lower panel shows the size of the convective
core (fractional mass) as the star evolves. The rapid change in the stellar
radius and the mass of the convective core at about 24.5 Myrs is the 
because the star is evolving off the main sequence.
}
\end{figure}

\begin{figure}     
\centerline{
\psfig{figure=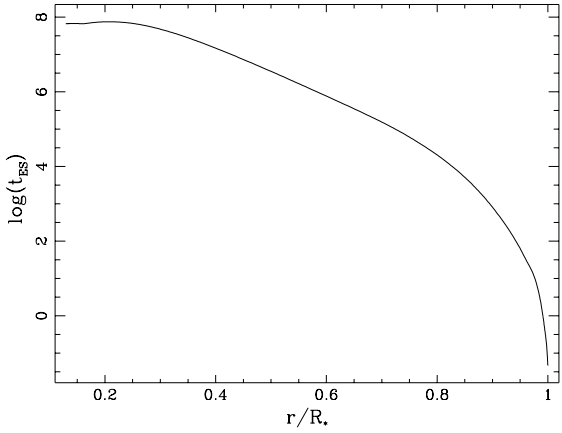,height=6cm,angle=-90}
}
\caption[]{Local
Local Eddington-Sweet timescale at radius $r$, $t_{ES}(r) = (GM_r^2/L_r) (4\pi
G\rho(r)/3\O^2)$, characterizing the adjustment of the rotation
profile through meridional circulation (in years) for a 9.0 $M_\odot$ main
sequence star of age 19.7 Myrs.
}
\end{figure}

\begin{figure}     
\centerline{
\psfig{figure=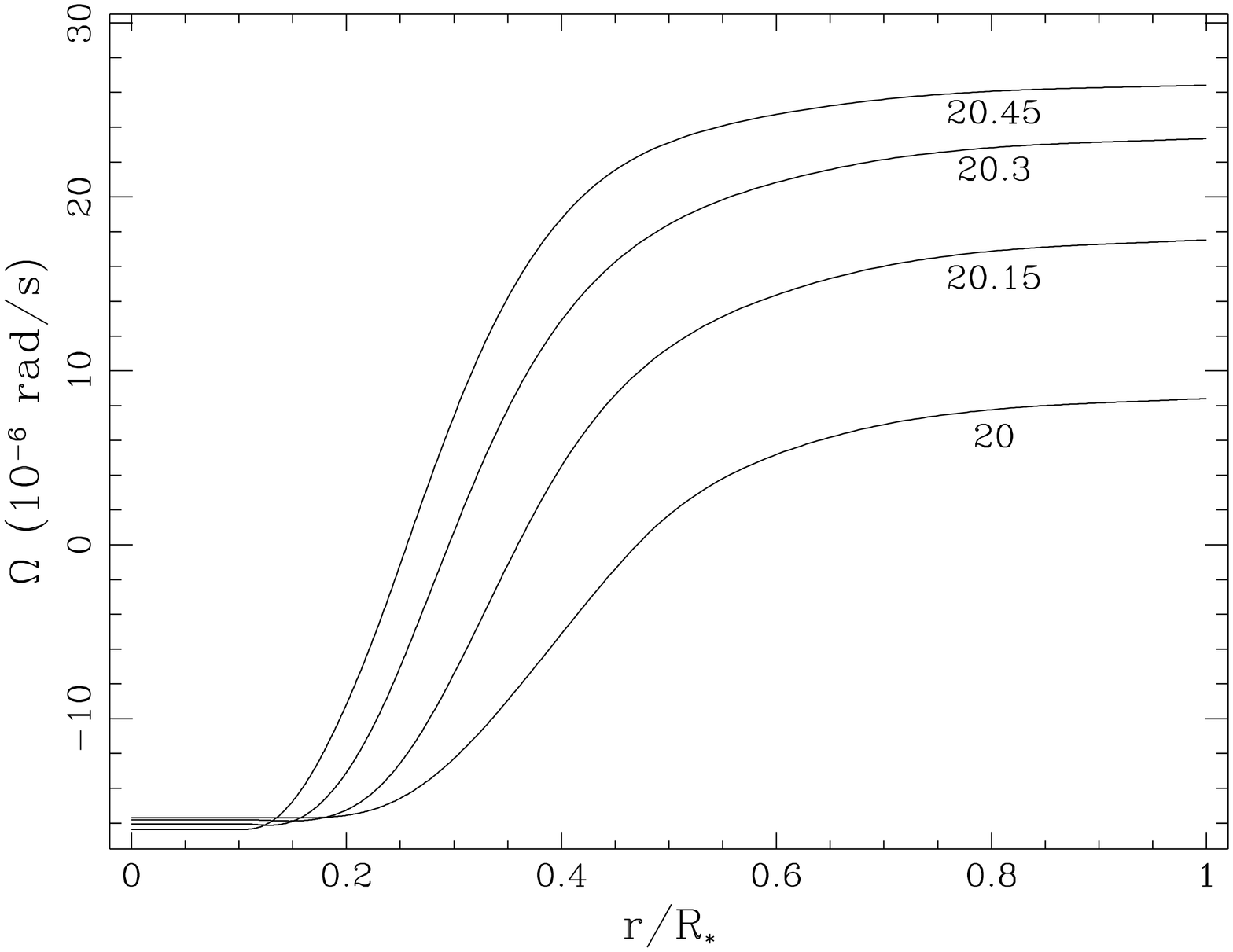,height=6cm,angle=-90}
}
\centerline{ 
\psfig{figure=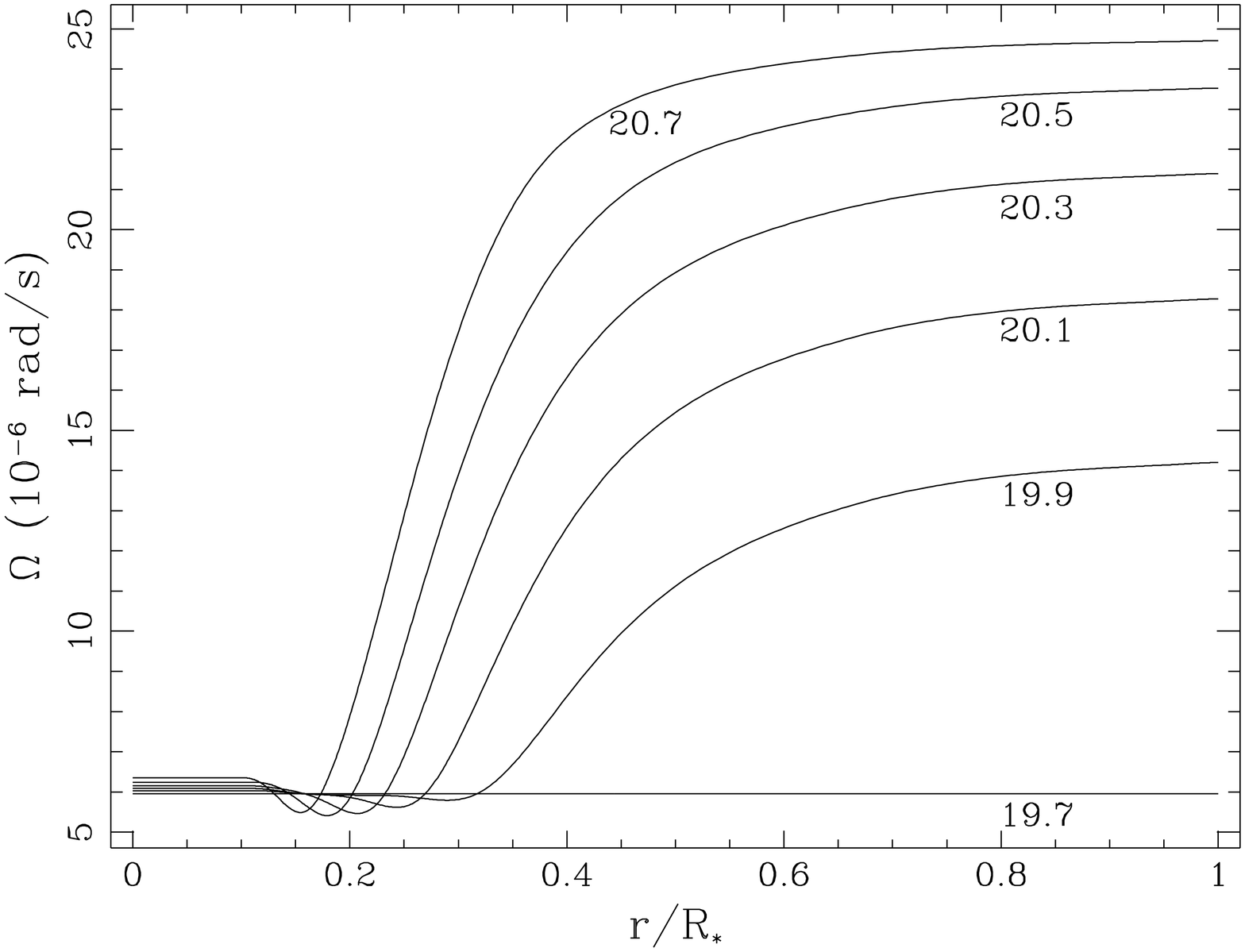,height=6cm,angle=-90}
}
\caption[]{
Rotation profile of the star at several times
(in millions of years). The top panel shows the case where the initial
rotation rate of the star was retrograde with angular speed of
$1.6 \times 10^{-5}$ rad s$^{-1}$, and the lower panel corresponds to the
prograde rotation speed of $6 \times 10^{-6}$ rad s$^{-1}$.
}
\end{figure}

\begin{figure}     
\centerline{
\psfig{figure=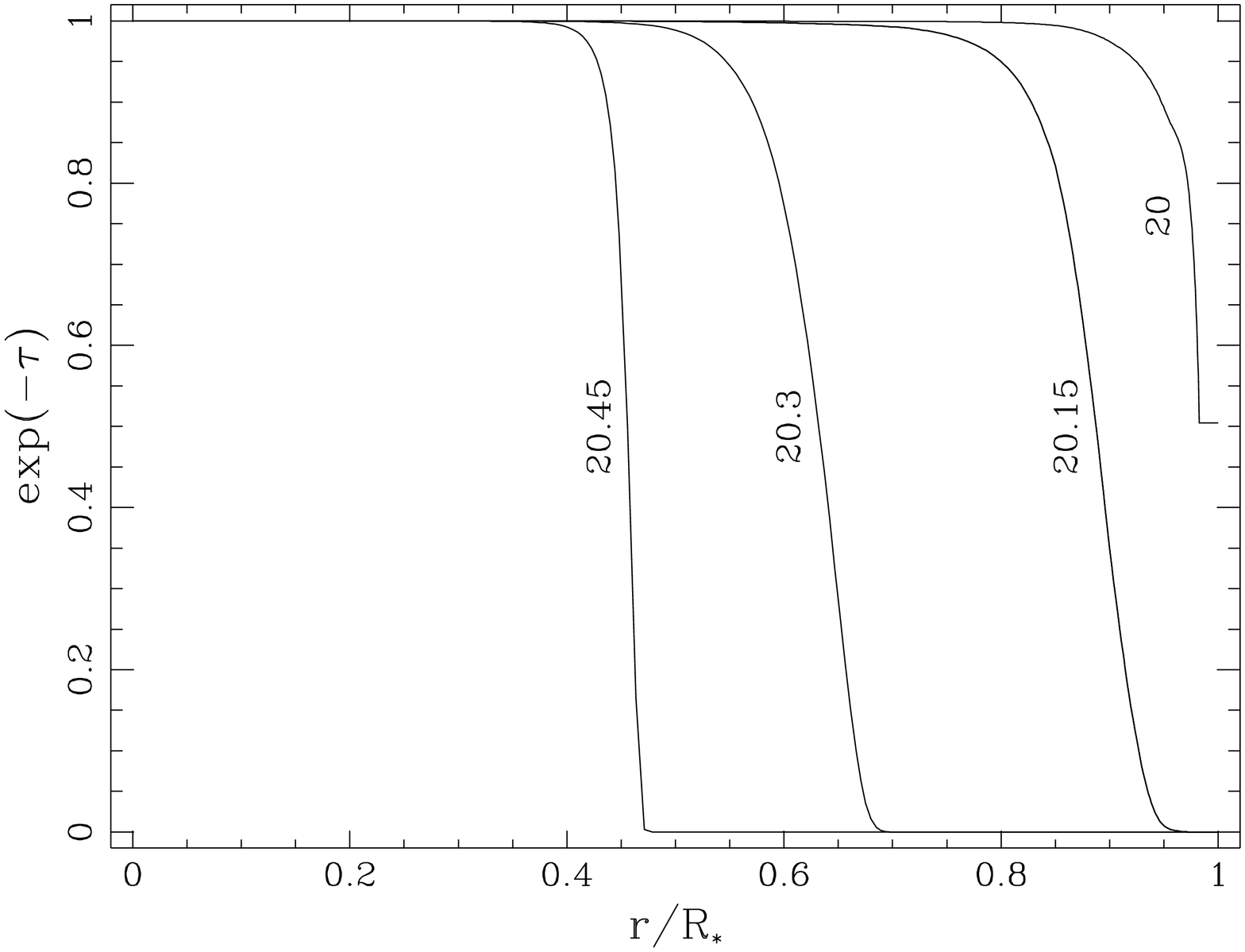,height=6cm,angle=-90}
}
\centerline{
\psfig{figure=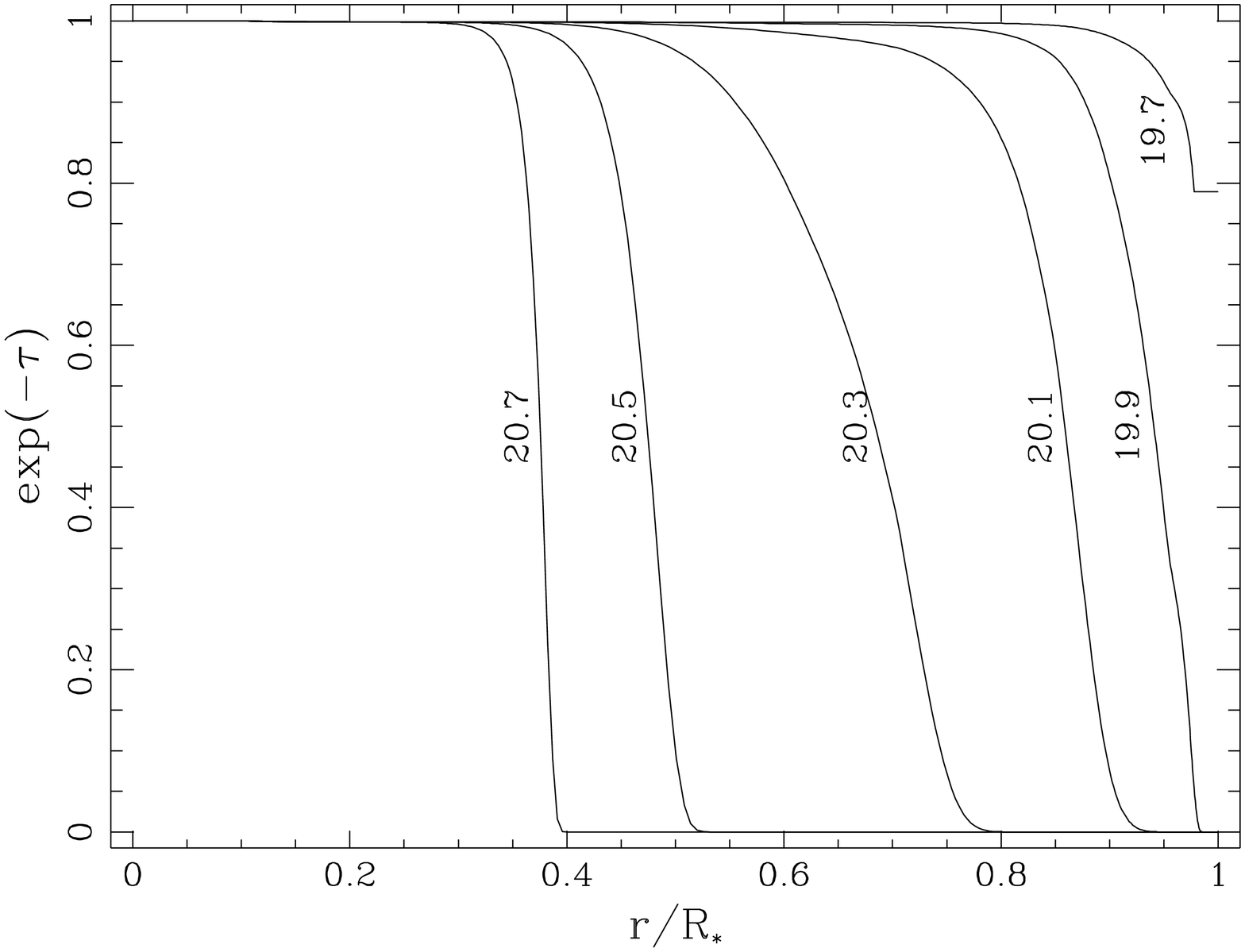,height=6cm,angle=-90}
}
\caption[]{
Relative magnitude of the momentum luminosity ${\cal L}(r)/{\cal L}(r_c)
= \exp (-\tau)$
as a function of time for the two
different cases of initial rotation of the B-star: retrograde rotation
with angular speed $1.6 \times 10^{-5}$ rad s$^{-1}$, and prograde rotation
rate of $6 \times 10^{-6}$ rad s$^{-1}$.
Note that with increasing time, as the rotation profile evolves and the 
thinkness of the synchronously rotating shell near the surface increases, 
the wave dissipation occurs at smaller $r$.
}
\end{figure}

\end{document}